\documentstyle[aps,prd]{revtex}
\date{\today}
\tighten
\begin{document}
\tighten

%%%%%%%%%%%%%%%%%%%%%%%%%%%%%%%%%%%%%%%%%%%%%%%%%%%%%%%%%%%%%%%%%%%%

\title{Gravitational Wave Detector Sites}
\author{B. Allen}
\address{Department of Physics, University of Wisconsin - Milwaukee,
P.O. Box 413, Milwaukee, Wisconsin 53201, U.S.A.}
\maketitle
\begin{abstract}
Locations and orientations of current and proposed
laser-interferometric gravitational wave detectors are given in tabular form.
\end{abstract}

%%%%%%%%%%%%%%%%%%%%%%%%%%%%%%%%%%%%%%%%%%%%%%%%%%%%%%%%%%%%%%%%%%%%
% Introduction
%%%%%%%%%%%%%%%%%%%%%%%%%%%%%%%%%%%%%%%%%%%%%%%%%%%%%%%%%%%%%%%%%%%%

\section{Introduction}
Gravitational waves are one of the most robust predictions of
Einstein's general theory of relativity, but have only been observed
indirectly, as the dominant energy-loss mechanism in the binary pulsar
PSR1913+16.  A new generation of laser-interferometric gravitational
wave detectors is currently under construction, which should permit
direct observations of these waves.  The analysis of signals from these
detectors, and the pioneering work on data analysis from existing
``prototype" detectors, requires correlating signals from different
sites in order to extract the most information.  Such analysis requires
precise knowledge of the locations and orientations of the detectors.
This paper presents a table of this data for the current and proposed
earth-based laser-interferometric detectors.

%%%%%%%%%%%%%%%%%%%%%%%%%%%%%%%%%%%%%%%%%%%%%%%%%%%%%%%%%%%%%%%%%%%%
% Format
%%%%%%%%%%%%%%%%%%%%%%%%%%%%%%%%%%%%%%%%%%%%%%%%%%%%%%%%%%%%%%%%%%%%

\section{Site and Orientation Data}
Each line of Table~\ref{table1} contains information about one
detector.  The data contained in each column is:
\begin{enumerate}
\item The name of the detector or project.
\item The name of the geographical site or location.
\item A nominal ``date of operation" (either past, current, or anticipated).
\item The arm length, in meters.
\item The location of the central (corner) station on the earth's
surface.  The lattitude is measured in degrees North from the equator,
and the longitude is measured in degrees West of Greenwich, England.
\item The orientation of the first arm, measured in degrees
counter-clockwise from true North.
\item The orientation of the second arm, measured in degrees
counter-clockwise from true North.
\item The source of information about the location and orientation.
\end{enumerate}

%%%%%%%%%%%%%%%%%%%%%%%%%%%%%%%%%%%%%%%%%%%%%%%%%%%%%%%%%%%%%%%%%%%%
% Notes
%%%%%%%%%%%%%%%%%%%%%%%%%%%%%%%%%%%%%%%%%%%%%%%%%%%%%%%%%%%%%%%%%%%%

\section{Notes}
\begin{enumerate}
\item The orientation of the Glasgow detector was changed in
1995, hence both the earlier and current orientations are given. 
\item The LIGO site in Hanford
Washington will have both 2 km and 4 km arms contained in the same vacuum tube.
\item The orientation
angles given for the GEO-600 site are {\it not} in error; for practical reasons
the arms are separated by $94.33^\circ$.
\end{enumerate}

%%%%%%%%%%%%%%%%%%%%%%%%%%%%%%%%%%%%%%%%%%%%%%%%%%%%%%%%%%%%%%%%%%%%
% Acknowledgements
%%%%%%%%%%%%%%%%%%%%%%%%%%%%%%%%%%%%%%%%%%%%%%%%%%%%%%%%%%%%%%%%%%%%
\acknowledgements

The author thanks participants in these different projects
for providing and verifying this site information. The work of BA was supported by NSF grants
PHY91-05935 and PHY95-07740.

%%%%%%%%%%%%%%%%%%%%%%%%%%%%%%%%%%%%%%%%%%%%%%%%%%%%%%%%%%%%%%%%%%%%
% References
%%%%%%%%%%%%%%%%%%%%%%%%%%%%%%%%%%%%%%%%%%%%%%%%%%%%%%%%%%%%%%%%%%%%

\begin{table}
\caption{ Site and orientation of earth-based interferometric
gravitational-wave detectors.}
\begin{center}
\begin{tabular}{|c|c|c|c|c|c|c|c|} \hline
{\sl Project}&{\sl Location}&{\sl Year}&{\sl Length} (m)&{\sl Corner
Location}&{\sl Arm 1}&{\sl Arm 2}&{\sl Source}\\ \hline
Glasgow & Glasgow, GBR & 1977 & 10 & $55.87^\circ$N $\quad
4.28^\circ$W  & $77.0^\circ$ & $167.0^\circ$ & \cite{Ruediger}
\\ \hline
CIT & Pasadena, CA, USA & 1980 & 40 & $34.17^\circ$N $\quad
118.13^\circ$W  & $180.0^\circ$ & $270.0^\circ$ & \cite{Raab} \\ \hline
MPQ & Garching, GER & 1983 & 30 & $48.24^\circ$N $\quad -11.68^\circ$W
& $329^\circ$ & $239^\circ$ & \cite{Ruediger} \\ \hline
ISAS-100 & Tokyo, JPN & 1986 & 100 & $35.57^\circ$N $\quad
-139.47^\circ$W  & $42.0^\circ$ & $135.0^\circ$ & \cite{Mizuno}
\\ \hline
TAMA-20 & Tokyo, JPN & 1991 & 20 & $35.68^\circ$N $\quad
-139.54^\circ$W  & $45.0^\circ$ & $315.0^\circ$ & \cite{Fujimoto}
\\ \hline
Glasgow & Glasgow, GBR & 1995 & 10 & $55.87^\circ$N $\quad
4.28^\circ$W  & $62.0^\circ$ & $152.0^\circ$ & \cite{Hough} \\ \hline
TAMA-300 & Tokyo, JPN & 1998 & 300 & $35.68^\circ$N $\quad
-139.54^\circ$W  & $90.0^\circ$ & $180.0^\circ$ & \cite{Fujimoto}
\\ \hline
GEO-600 & Hannover, GER & 1999 & 600 & $52.25^\circ$N $\quad
-9.81^\circ$W  & $25.94^\circ$ & $291.61^\circ$ & \cite{Ruediger}
\\ \hline
VIRGO   & Pisa, ITA & 2000  & 3000        & $43.63^\circ$N $\quad
-10.5^\circ$W  & $71.5^\circ$      & $341.5^\circ$     &
\cite{Poggiani} \\ \hline
LIGO    & Hanford, WA, USA  & 2000  & 4000        & $46.45^\circ$N
$\quad 119.41^\circ$W & $36.8^\circ$      & $126.8^\circ$  &
\cite{Raab} \\ \hline
LIGO    & Livingston, LA, USA & 2000  & 4000        & $30.56^\circ$N
$\quad 90.77^\circ$W  & $108.0^\circ$     & $198.0^\circ$ & \cite{Raab}
\\ \hline
\end{tabular}
\end{center}
\label{table1}
\end{table}

\end{document}